\documentstyle[12pt]{article}
\begin{document}
\begin{flushright}
IHEP 95-80\\
hep-ph/9507228\\
\end{flushright}
\vspace*{4mm}
\begin{center}
{\large \bf Hard-soft factorization in $B_c^+\to \psi \pi^+$ decay}\\
\vspace*{4mm}
V.V.Kiselev\\
Institute for High Energy Physics,\\
Protvino, Moscow Region, 142284, Russia.\\
E-mail: kiselev@mx.ihep.su\\
Fax: +7-(095)-230 23 37.
\end{center}
\begin{abstract}
The width of $B_c^+\to \psi \pi^+$ decay is calculated in the framework of
factorization of a hard gluon exchange and a soft binding of quarks in
the heavy quarkonium, so that ${\rm BR}(B_c^+\to \psi \pi^+) = 2.0\pm 0.5$ \%.
\end{abstract}

\section{Introduction}
In the recent review on the $B_c$ meson physics \cite{ufn}, one analyzed
possibilities of a practical observation of the long-living pseudoscalar
$1S$-state in the flavoured heavy quarkonium $(\bar b c)$ at running
facilities. One shown, that at present integral luminocities of the LEP and
FNAL experiments, the expected number of completely reconstructed events
with the weak decays of $B_c$ is of the order of one or two dozens, where
the specific place is taken by the $B_c$ decay modes with the $\psi$
particle in the final state, since the latter can be observed with a high
efficiency in the leptonic mode of $\psi\to l^+l^-$. In contrast to the
manyparticle or semileptonic  modes of decay, the $B_c^+\to \psi \pi^+$
transition with $\psi\to l^+l^-$ is outstanding, because one can completely
reconstruct the $B_c$ decay vertex over the tracks by the charged particles
and one can measure the $B_c$ mass. Therefore, the value of branching ratio
for the $B_c^+\to \psi \pi^+$ decay is considered to be quite significant for
the practical search of $B_c$.

At present, estimates of the $B_c^+\to \psi \pi^+$ decay width have been done
in the framework of nonrelativistic potential models for constituent
quarks \cite{kt,eg,dw,lus,ch} with account for hard gluon corrections to the
weak transition currents as $b\to cW^{-\ast}$. The different models give
the branching ratio value, close to
$$
{\rm BR}^{\rm PM}(B_c^+\to \psi \pi^+)\approx 0.2 \%\;.
$$
However, the application of such potential models is assumed to have the
maximum accuracy at low momenta of the recoil meson ($\psi$ in the decay
under consideration), since the transition form factors are determined by
integrals of overlapping between wave functions of the decaying meson and
the recoil one over the momentum region, which is the most accurately
determined at low recoils, when maxima of the wave functions overlapp, and
these maxima are the most certainly calculated in the nonrelativistic approach.
At large recoil momenta, it is necessary to calculate the overlapping between
the maximum of the first function and the rapidly decreasing tail of the
second wave function\footnote{One usually takes the oscillator wave functions
as the model ones, so that the decrease with the rise of momentum squared
is exponential.}, so that the procedure leads to a large numerical uncertainty,
having a principial sense. Indeed, in the framework of the nonrelativistic
approximation, the meson is considered as a narrow wave packedge of free quarks
with a low binding energy. Therefore, this description is applicable only
at low relative momenta between the constituent quarks, and the model is
selfconsistent only under the condition, that the wave functions, obtained
in such way, give a suppression of large relative momenta. For the description
of the latters, the nonrelativistic potential models do not claim, since
a contribution of this region, say, into the Schr\"odinger equation is
suppressed by the sense of the potential model construction. One usually
supposes, that a momentum, being of the order of the constituent quark mass,
is the low boundary of large relative momenta. As one can easily show
in the $B_c^+\to \psi \pi^+$ decay, the situation takes place, when the
application of the wave function overlapping formalism is not excusive and it
can serve for the exptrapolation estimates in the order of magnitude\footnote{
Moreover, in ref.\cite{cw} one shown, that a correct covariant description of
form factors by the wave function overlapping, results in an additional
factor, enforcing the suppression of large recoil momenta, so that the
extrapolation estimate of $B_c^+\to \psi \pi^+$ width must be made to a lower
value.}. In this case, one has to consider the quark-meson form factor in the
region, where one of the quarks, entering the meson, has a large virtuality.
Hence, this region can not be described in the framework of nonrelativistic
form factors, where, by construction, both constituent quarks are in vicinity
of the mass shells. However, the heavy quark virtualities, comparable with
its mass, correspond to the field of applicability of the QCD perturbation
theory. Therefore, one can consider a hard gluon emission by the low
virtuality quark, that can be described in the framework of the potential
approach. Thus, we can factorize the amplitude of the weak decay of
heavy quark with account for the hard gluon exchange with the spectator quark
and the amplitude of soft binding of "free" quarks into the heavy quarkonium.

In the present paper, in the framework of this hard-soft factorization, we
perform the calculation of the $B_c^+\to \psi \pi^+$ decay width under the
diagrams on figure 1, so that
$$
{\rm BR}(B_c^+\to \psi \pi^+) = 2.0\pm 0.5 \%\;,
$$
that is quite greater than the extrapolation estimates of potential models.
Therefore, we believe, that the possible probability of the practical
$B_c$ observation in the current experiments is enforced in the
$B_c^+\to \psi \pi^+$ mode.

\section{Calculation of $B_c^+\to \psi \pi^+$ width}

In the framework of the nonrelativistic formalism for the heavy quark binding
into the $S$-wave quarkonium, we assume, that the quark momentum, entering the
meson, is equal to $p_Q^\mu = m_Q v^\mu$, where $v_\mu$ is the four-velocity of
quarkonium, so that the quarks inside the meson move with one and the same
four-velocity $v$. Moreover, the quark-meson vertex with nontruncated
quark lines corresponds to the spinor matrix
$$
\Gamma_V = \hat \epsilon\; \frac{1+\hat v}{2}\;
\frac{\tilde f M_{nS}}{2\sqrt{3}}\;,
$$
for the vector quarkonium with $\epsilon_\mu$, being the polarization vector,
and
$$
\Gamma_P = \gamma_5\; \frac{1+\hat v}{2}\;
\frac{\tilde f M_{nS}}{2\sqrt{3}}\;,
$$
for the pseudoscalar quarkonium, so that $M_{nS}$ is the $nS$-level mass and
$\tilde f$ is related with the value of configuration wave function at origin
$$
\tilde f = \sqrt{\frac{12}{M_{nS}}}\; |\Psi_{nS}(0)|\;.
$$
The $\tilde f$ quantity can be related with the leptonic constants of states
\begin{eqnarray}
\langle 0| J_\mu(0)|V\rangle & = & i f_V M_V\; \epsilon_\mu\;, \nonumber\\
\langle 0| J_{5\mu}(0)|P\rangle & = & i f_P p_\mu\;, \nonumber
\end{eqnarray}
where $J_{\mu}(x)$ and $J_{5\mu}(x)$ are the vector and axial-vector currents
of the constituent quarks. Then the account for hard gluon corrections
in the first order over $\alpha_s$ \cite{sv,bra,vol,kis} results in
\begin{eqnarray}
\tilde f & = & f_V\; \bigg[1 - \frac{\alpha_s^H}{\pi}
\biggl(\frac{m_2-m_1}{m_2+m_1}\ln\frac{m_2}{m_1} -\frac{8}{3}\biggr)\bigg]\;,\\
\tilde f & = & f_P\; \bigg[1 - \frac{\alpha_s^H}{\pi}
\biggl(\frac{m_2-m_1}{m_2+m_1}\ln\frac{m_2}{m_1} - 2\biggr)\bigg]\;,
\end{eqnarray}
where $m_{1,2}$ are the masses of quarks, composing the quarkonium. For the
vector currents of quarks with equal masses, the BLM procedure of the scale
fixing in the "running" coupling constant of QCD \cite{blm} gives \cite{vol}
$$
\alpha_s^H = \alpha_s^{\overline{\rm MS}}(e^{-11/12}m_Q^2)\;.
$$
For the quarkonium with $m_1\neq m_2$, we assume
$$
\alpha_s^H = \alpha_s^{\overline{\rm MS}}(e^{-11/12}m_1 m_2)\;.
$$
Further, the factor of the colour wave function $\delta^{ij}/\sqrt{3}$ stands
in the quark-meson vertex. The $\pi$ meson current corresponds to the
axial-vector current of weak transition $A^\mu =f_\pi p_\pi^\mu$.

Then the matrix element of $B_c^+\to \psi \pi^+$ decay, calculated
under the diagrams on figure 1, takes the form
\begin{equation}
T(B_c^+\to \psi \pi^+) = \frac{G_F}{\sqrt{2}}\; V_{bc}\; \frac{32\pi\alpha_s}
{9}\; f_\pi \tilde f_{B_c}\tilde f_{\psi}\; \frac{M^2}
{m_\psi^2 (y-1)^2}\; (\epsilon \cdot v)\;,
\label{t}
\end{equation}
where $v$ is the four-velocity of $B_c$ meson, $M$ is its mass, $\epsilon$
is the polarization vector of $\psi$ particle, $m_\psi$ is its mass,
$y=v\cdot v_\psi$ is the product of the $B_c$ and $\psi$ four-velocities,
$$
y=\frac{M^2+m^2_\psi}{2M m_\psi}\;.
$$
In eq.(\ref{t}) $\alpha_s$ is given at the scale of the gluon virtuality
\begin{equation}
k^2_g = - m^2_\psi (y-1)/2 \approx -1.2\; \mbox{GeV}^2\;,
\label{kg}
\end{equation}
so that
$$
\alpha_s = \alpha_s^{\overline{\rm MS}}(-e^{-5/3}k^2_g)\;.
$$
As is seen from eq.(\ref{kg}), the virtuality of hard gluon is comparable
with the square of charmed quark mass, and it points to the applicability
of the hard process factorization. The corresponding virtualities of $\bar c$
and $\bar b$ quarks, interacting with the hard gluon, are equal to
\begin{eqnarray}
k^2_c - m^2_c & = & 2k^2_g\;, \nonumber\\
k^2_b - m^2_b & = & 2 k^2_g M/m_\psi\;. \nonumber
\end{eqnarray}
Note, that one numerically gets, that the virtual $\bar c$-quark in the second
diagram of figure 1 is in the $t$-channel, since its four-momentum has a
negative square. Therefore, one can see, that the corresponding contribution
into the $B_c^+ \to \psi \pi^+$ decay is quite a nonspectator one.

{}From eq.(\ref{t}) one gets the expression for the total width of
$B_c^+\to \psi \pi^+$ decay
\begin{equation}
\Gamma(B_c^+\to \psi \pi^+) = G_F^2 |V_{bc}|^2\; \frac{128\pi\alpha_s^2}{81}
f^2_\pi \tilde f^2_{B_c}\tilde f^2_{\psi} \biggl(\frac{M+m_\psi}{M-m_\psi}
\biggr)^3\; \frac{M^3}{(M-m_\psi)^2 m_\psi^2}\;. \label{g}
\end{equation}
In numerical estimates, we suppose \cite{neu}
$$
|V_{bc}| = 0.041\pm 0.003\;,
$$
and we use the one-loop expression for the $\alpha_s$ evolution
$$
\alpha_s(m^2) = \frac{4\pi}{\beta_0(n_f)\ln(m^2/\Lambda^2_{(n_f)})}\;,
$$
where $\beta_0(n_f)=11-2n_f/3$, $n_f$ is the number of quark flavours with
$m_{n_f}<m$,
$$
\Lambda_{(n_f)} = \Lambda_{(n_f+1)} \biggl(\frac{m_{n_f+1}}{\Lambda_{(n_f+1)}}
\biggr)^{\frac{2}{3\beta_0(n_f)}}\;.
$$
Using $\alpha_s^{\overline{\rm MS}}(m_Z^2) = 0.117\pm 0.005$ \cite{pdg},
one finds, that $\Lambda_{(5)} = 85\pm 25$ MeV and $\Lambda_{(3)} =
140\pm 40$ MeV.

Further, the $f_{B_c}$ constant was estimated in the framework of QCD sum rules
\cite{kis,nar,np}
$$
f_{B_c} = 385\pm 25\; \mbox{MeV,}
$$
and it is in a good agreement with the scaling relation for the leptonic
constants of $1S$ heavy quarkonia \cite{np,kis}
$$
\frac{f^2}{M}\; \biggl(\frac{M}{\mu_{12}}\biggr)^2 = {\rm const.}\;,\;\;\;\;\;
\mu_{12} = \frac{m_1 m_2}{m_1+m_2}\;.
$$
Then the account for the hard gluon corrections gives
\begin{eqnarray}
\tilde f_\psi & = & 542\pm 50\; \mbox{MeV,} \\
\tilde f_{B_c} & = & 440\pm 40\; \mbox{MeV.}
\end{eqnarray}

The total width of the decay is equal to
\begin{equation}
\Gamma(B_c^+\to \psi \pi^+) = (24\pm 5)\cdot 10^{-6}\; {\rm eV} =
\frac{1}{26\pm 5\; {\rm ps}}\;, \label{gam}
\end{equation}
where we have supposed $M= 6.25$ GeV \cite{ufn}.

To calculate the branching ratio of the $B_c^+\to \psi \pi^+$ decay, we
evaluate the total $B_c$ meson width due to the formula \cite{ufn,lus}
$$
\Gamma(B_c) \approx \Gamma(B) + (0.6\pm 0.1) \Gamma(D^+) + \Gamma(ann.)\;,
$$
where $\Gamma(B)$ is the contribution of $\bar b$-quark decays with
the $c$-quark, being the spectator, $\Gamma(D^+)$ determines the contribution
of $c$-quark decays with the $\bar b$-quark, being the spectator, and with
the account for the phase space reduction, because of the $c$-quark binding
inside $B_c$ (i.e. one takes into account the deviation from the exact
spectator consideration), and $\Gamma(ann.)$ is the contribution of
annihilation channels, depending on $|V_{bc}|$ and $f_{B_c}$. Then
$$
\Gamma(B_c) = (1.2\pm 0.2)\cdot 10^{-3}\; {\rm eV} = \frac{1}{0.55\pm 0.1\;
{\rm ps}}
$$
and
\begin{equation}
{\rm BR}^{\rm HS}(B_c^+\to \psi \pi^+) = 2.0\pm 0.5 \%\;. \label{br}
\end{equation}
Note, that in the given estimates of eqs.(\ref{gam}) and (\ref{br}), we
do not include into the consideration the $a_1$ factor, caused by the
hard gluon corrections to the effective four-fermion weak interactions of
quarks. $a_1$ is evaluated for free quarks, so that $a_1= 1.22\pm 0.04$
\cite{ufn}. In the current consideration, one of the heavy quarks in the
$\bar b\to \bar c W^{+\ast}$ transition is hardly virtual, and we use
the factorization of $\pi$ meson current, so that the corresponding $\alpha_s$
correction would look as the higher order gluon contribution to the
$W$-boson and quark vertex. We do not account for such higher order
corrections in the present consideration.

Further, the purely spectator decay of $\bar b \to \bar c \pi^+$ has the total
width, equal to
$$
\Gamma(\bar b \to \bar c \pi^+) = G_F^2 |V_{bc}|^2\; \frac{m_b^3 f_\pi^2}
{16\pi}\; \biggl(1- \frac{m_c^2}{m_b^2}\biggr)^3\; a_1^2\;,
$$
that corresponds to
\begin{equation}
{\rm BR}^{\rm spec}(\bar b \to \bar c \pi^+)\approx 0.8\%\;, \label{spec}
\end{equation}
where the axial-vector current contribution is equal to one half of estimate
(\ref{spec}). On the other hand, the matrix element, corresponding to
the first diagram on figure 1, approximately equals the matrix element,
following from the second diagram, and hence, estimate (\ref{br}) is four
times enforced due to the $t$-exchange nonspectator contribution. We
believe, that the given enhancement is quite acceptable.

\section{Conclusion}

In the present paper we have shown, that in the $B_c^+\to \psi \pi^+$ decay,
the hard recoil momentum of $\psi$ particle leads to that, the formalism
of the weak transition form factor calculation, based on the overlapping
between the nonrelativistic wave functions for the heavy quarkonia, goes
out the framework of the applicability, since the hard gluon exchange with the
spectator quark results in the large virtuality of heavy quark in the
weak transition current. In the considered case, the amplitude of the weak
decay with the hard exchange by gluon can be calculated in the framework of
QCD perturbation theory and it can be factorized from the amplitude of soft
binding of heavy quarks in the quarkonium. The calculation with the account
for this hard-soft factorization results in
$$
{\rm BR}(B_c^+\to \psi \pi^+) = 2.0\pm 0.5 \%\;,
$$
where the accuracy is basically restricted by uncertainties in the evolution
scale of the QCD "running" coupling constant, in the charmed quark mass and
the total $B_c$ width. The given estimate of the branching ratio for the
$B_c^+\to \psi \pi^+$ decay mode is significantly greater than the
extrapolation results of potential models. This value strongly enforces a
practical probability of $B_c$ observation in the current FNAL and LEP
experiments with vertex detectors.

The author thahks A.Razumov and S.Slabospitsky for the software help in
the figure creation.

This work is partially supported by the International Science Foundation
grants NJQ000, NJQ300 and by the program "Russian State Stipendia for young
scientists".

\vspace*{4mm}
\hfill {\it Received June 9, 1995}
\end{document}